\documentclass[a4paper,10pt,preprint,5p,number, lefttitle]{elsarticle}

\usepackage{hyperref}
\usepackage{xspace}
\newcommand{\kHz}{\ensuremath{\mathrm{kHz}}\xspace}
\newcommand{\MHz}{\ensuremath{\mathrm{MHz}}\xspace}
\newcommand{\mm}{\ensuremath{\mathrm{mm}}\xspace}
\newcommand{\Ar}{\ensuremath{\mathrm{Ar}}\xspace}
\newcommand{\COtwo}{\ensuremath{\mathrm{CO}_2}\xspace}
\newcommand{\GB}{\ensuremath{\mathrm{GB}}\xspace}
\newcommand{\GBs}{\ensuremath{\mathrm{GB/s}}\xspace}
\newcommand{\MBs}{\ensuremath{\mathrm{MB/s}}\xspace}
\newcommand{\Gbps}{\ensuremath{\mathrm{Gbps}}\xspace}

\makeatletter
\def\ps@pprintTitle{%
 \let\@oddhead\@empty
 \let\@evenhead\@empty
 \def\@oddfoot{}%
 \let\@evenfoot\@oddfoot}
\makeatother

\begin{document}

\begin{frontmatter}

\title{A horizontally scalable online processing system for trigger-less data acquisition}

\author[1,2]{Matteo Migliorini}
\author[1,2]{Jacopo Pazzini\corref{cor}}
\ead{jacopo.pazzini@unipd.it}
\author[1,2]{Andrea Triossi}
\author[1,2]{Marco Zanetti}
\author[2]{Alberto Zucchetta}

\address[1]{Department of Physics and Astronomy “Galileo Galilei”, Padova University, Via Marzolo 8, 35131 Padova, Italy}
\address[2]{National Institute for Nuclear Physics, Padova Division, Via Marzolo 8, 35131 Padova, Italy}

\cortext[cor]{Corresponding author}

\begin{abstract}
\noindent
The vast majority of high energy physics experiments rely on data acquisition and hardware-based trigger systems performing a number of stringent selections before storing data for offline analysis. 
The online reconstruction and selection performed at the trigger level are bound to the synchronous nature of the data acquisition system, resulting in a trade-off between the amount of data collected and the complexity of the online reconstruction performed.
Exotic physics processes, such as long-lived and slow-moving particles, are rarely targeted by online triggers as they require complex and nonstandard online reconstruction, usually incompatible with the time constraints of most data acquisition systems.
The online trigger selection can thus impact as one of the main limiting factors to the experimental reach for exotic signatures.
Alternative data acquisition solutions based on the continuous and asynchronous processing of the stream of data from the detectors are therefore foreseeable as a way to extend the experimental physics reach.
Trigger-less data readout systems, paired with efficient streaming data processing solutions, can provide a viable alternative.
In this document, an end-to-end implementation of a fully trigger-less data acquisition and online data processing system is discussed. 
An easily scalable and deployable implementation of such an architecture is proposed, based on open-source distributed computing frameworks capable of performing asynchronous online processing of streaming data.
The proposed schema can be suitable for deployment as a fully integrated data acquisition system for small-scale experimental apparatus, or to complement the trigger-based data acquisition systems of larger experiments.
A muon telescope setup consisting of a set of gaseous detectors is used as the experimental development testbed in this work, and a fully integrated online processing pipeline deployed on cloud computing resources is implemented and described.
%
\end{abstract}


\end{frontmatter}


\section{Introduction}\label{sec:Introduction}

The data acquisition (DAQ) and trigger systems are the cornerstone components of any modern physics experiment, defining the path and establishing the bandwidth at which the data produced in the experimental apparatus is digitized, filtered, and permanently stored for offline processing and analysis.
In particle physics experiments, multi-staged DAQ and trigger architectures are usually implemented~\cite{cms_daq,CMS:2016ngn,Bawej:2014atd, atlas_daq, ATLASTDAQ:2016pov, lhcb_daq} to cope with the large data throughput produced by the sensing elements.
All experiments' sub-detectors are digitized and readout synchronously based on a common clock distributed across the entire experiment.
A number of online trigger stages implement sets of simple reconstruction and selection criteria.
A first trigger stage is usually based on custom electronic boards (typically hosting Field Programmable Gate Arrays chips, FPGAs) which filter the data stream lowering the throughput to the level at which a further fully software-based processing stage can be manageable.
This DAQ paradigm embodies a trade-off between the experiments' goal to maximize the amount of data collected and the technological limitations in online processing and storing data capabilities.
For example, in a typical large collider experiment as the Compact Muon Solenoid (CMS)~\cite{cms}, the first level hardware-based trigger reduces the data rate from the 40~\MHz collision rate (also referred to as bunch crossing rate) to 100~\kHz.
Due to the synchronous nature of the CMS trigger, the time required to take the first trigger level decision is limited to a few microseconds.
Data is hosted in dedicated buffers waiting for the hardware trigger decision which either accepts or rejects the entire event. 
If a trigger decision is not taken within this time window, the buffers are filled with new data, thus resulting in the loss of the oldest buffered event.
Accepted data are sent to a second software-based trigger stage, where the rate of events stored on disk is further reduced by two orders of magnitude.
All online selections are specifically designed to target the physics processes which represent the core of the experiments' scientific programs, and are defined by detailed studies and simulations of the Standard Model (SM) and a number of its proposed extensions.
Nevertheless, the physics potential of an experiment may be directly limited by the DAQ capability of processing and storing just a minor fraction of the produced data.
New and unexpected phenomena may be produced, but not collected due to the stringent trigger rules, based on a limited number of physics scenarios.
To extend the experiments' physics reach, a DAQ system should allow fast online analysis on a larger fraction of events, and possibly perform alternative reconstructions that may be the key to finding anomalies.
A noticeable example is the search for heavy stable charged particles (HSCP), exotic particles characterized by a long lifetime that traverse the experiment at slow velocity ($\beta \ll 1$)~\cite{Lee:2018pag}.
These particles are expected to release signals in the detector over a broad time window, spanning over multiple consecutive bunch crossings.
As the trigger system is synchronous with the bunch crossing rate of the collider, no online filter can easily take into account such late signals, delayed by several bunch crossings.
Dedicated trigger strategies can be implemented by introducing special adjustments to the hardware delays which affect the entire trigger chain.
Nonetheless, this approach has a sizable impact on the overall already stringent latency requirements of the trigger system, and cannot be generalized to account for arbitrary slow-moving particles.
Most HSCP searches currently performed at colliders thus rely on alternative triggers based, among others, on the presence of large missing transverse momentum or isolated energetic muons.
These trigger selections finally impact the reach of long-lived particle searches, hence limiting the experiments' discovery potential of HSCPs.
A thorough online analysis performed on unfiltered events may instead provide the collection and online processing of asynchronous data, thus reconstructing HSCP tracks spanning several bunch crossings, finally allowing to probe more extensively the phase space of the HSCP production.
 
In recent years, various experimental physics collaborations have implemented alternative DAQ architectures geared towards the exploitation of the available hardware and computing resources to increase the fraction of data collected and stored for offline analysis.
Noticeable examples are the extensive software-based reconstruction and calibration provided by the LHCb TurboStream~\cite{lhcb_turbo}, and the Xenon1T~\cite{xenon1t} DAQ system, centered around a fast NoSQL database to implement flexible software trigger strategies.

In the last decade, the increased availability of open-source frameworks for distributed data processing, in conjunction with the availability of more powerful computing resources, has made it possible to follow different paths in the implementation of DAQ systems.
In this work, a prototype for a novel technique for online data processing is presented.
In the proposed scheme, all data produced by the detectors are ideally processed asynchronously by means of fast distributed computing frameworks, making this DAQ scheme trigger-less in nature.
The system is envisioned to be able to cope with a wide range of input data throughput and to be easy to deploy in several computing contexts, both on on-premises and cloud-based, not relying on dedicated hardware or networking infrastructures.
For these reasons, it was chosen to build this processing scheme around modern distributed computing frameworks developed for big-data applications.
Such systems are designed to scale their performance with the number of computing resources available, a feature often referred to as horizontal scalability, and thus may be appropriately dimensioned to adapt to the number of input sources and event rates.
The computing resources are dynamically set up, configured, and managed via a resource orchestrator, Kubernetes~\cite{Kuberneteswebsite}, ensuring the portability of the proposed DAQ over any computing infrastructure.
A description of the first end-to-end implementation of the processing scheme is presented in this paper, where a set of muon detectors is used as a testbed for the development of the DAQ scheme and its first deployment.

Even in modern DAQ implementations only partially relying on hardware trigger selection, such as the TurboStream developed by the LHCb collaboration, custom computing and networking are required.
Data-fragments produced as the outcome of a single event by several sub-detectors have to be routed towards dedicated builder units, which perform the event reconstruction and eventually apply a software-based filtering stage.
The system described in this paper is instead centered around processing frameworks with no specific requirement on networking and computing resources, resulting in an easy-to-implement data acquisition system.
This is further achieved thanks to the centralized orchestration of all computing services provided by Kubernetes.

For these reasons, the proposed DAQ solution is expected to be applicable in a number of experimental contexts.
A potentially interesting application is the case of small-to-intermediate scale experimental apparatus employed in short data-taking campaigns, such as in the case of typical test-beams.
The design of a dedicated DAQ system integrating multiple independent detectors is usually a critical component, and the ad hoc definition of hardware-based triggers might result in a severely limiting factor in the quality and quantity of data collected.
A fully software online processing stage of a trigger-less data stream is instead expected to mitigate these issues, while at the same time providing live monitoring of all data collected by the detectors. 
This approach can be also thought to be integrated into larger-scale experiments with dedicated and highly specialized DAQ systems.
In view of the synchronous nature of the experiments' trigger and DAQ, the asynchronous processing schema presented in this paper may provide a complementary data path, able to target those processes (such as the HSCP) unable to be efficiently selected in the synchronous acquisition schema. 

\section{Experimental setup}\label{sec:setup}

\subsection{Drift Tube Detectors}

A set of gaseous detectors is used in the testbed experimental setup for the development of the proposed data acquisition scheme.
This set of detectors is used in a number of configurations ranging from a muon telescope setup to a dedicated tracking spectrometer for large fluence applications.
The detectors are based on the design of the Drift Tubes (DT)~\cite{CMS:2009lnj} of the Compact Muon Solenoid (CMS) experiment~\cite{cms}, and have been first employed to perform tracking and particle identification of muons during the 2019 test-beam campaigns of the LEMMA project~\cite{lemma}.
The wide range of throughput to which this set of detectors is exposed depending on their application makes for an excellent testbed for the development of a general-purpose DAQ processing schema.
It is however to be noted that this setup is only used as a testbed for the development of the trigger-less online processing system, and can be regarded as representative of a generic set of detectors.

Each detector, referred to as \textit{chamber}, is composed by 4 layers of 16 tubes (\textit{cells}), each with a transverse size of $42\times 13 \, \mm^2$. Cells of adjacent layers are mounted in a staggered configuration, with a relative shift of half the cell width, to provide tracking capabilities. 
As a charged particle passes through a cell, it induces ionization in the 85\%-15\% \Ar-\COtwo gas mixture.
The primary electrons drift towards the anodic wire, kept at positive potential and positioned at the center of the cell.
Close to the wire, secondary ionization occurs with a typical gain of $10^5$, and the electron avalanche is collected by the sensing wire.
The electric field inside the cell is appropriately shaped with a multi-electrode design to guarantee a constant drift velocity of the electron cloud throughout the entire cell.
Up to four time-coherent signals (\textit{hits}) can be produced by a muon crossing a chamber. 
The time coherence among the hits is due to the maximum time allowed for the signal in each cell to form and be collected by each wire, as well as the angle at which the muon crosses the tubes. 
The charge collected on each wire is amplified, shaped, and discriminated against a fixed threshold by an ASIC~\cite{cms_mad} hosted within the gas volume of the chamber. 
Hits passing the discriminator threshold are finally transmitted to the read-out electronics using the LVDS standard.

\subsection{Trigger-less Readout}

\begin{figure}[thb]
    \centering
    \includegraphics[width=\linewidth]{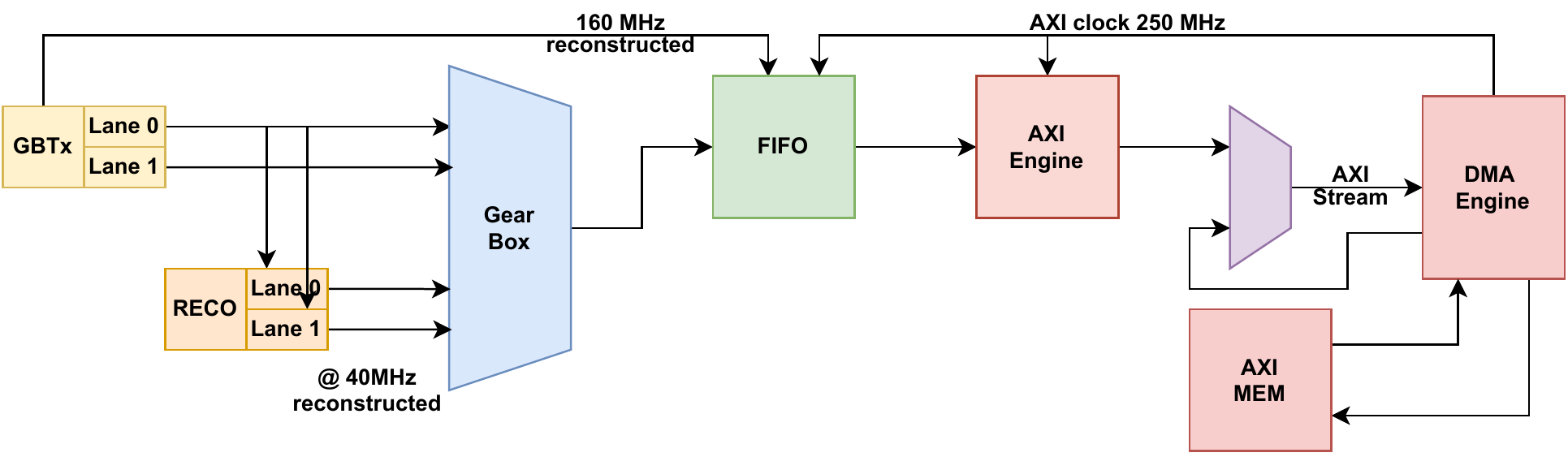}
    \caption{Schematic representation of the firmware implemented in the BE board.}
    \label{fig:kcufw}
\end{figure}

Two Xilinx VC707 evaluation boards, equipped with Virtex-7 XC7VX485T FPGAs, are used in this setup as front-end (FE) read-out boards.
Each board performs the time-to-digital conversion (TDC) of the LVDS signals to tag the time of arrival of the hits with respect to a reference clock.
The TDC is implemented in firmware (FW) using the standard IOSERDES running at 1.2 GSps.
An external oscillator is used to provide a 120~\MHz clock, distributed across both VC707 boards.
To emulate a typical LHC experiment clock distribution system~\cite{cms_tcds}, a 40~\MHz clock is generated internally to each VC707 by down-scaling the external reference clock. 
The TDC provides a time measurement of the signal rising edge with respect to the 40~\MHz clock. 
Each VC707 can collect a total of 138 channels, thus being able to collect hits from two chambers.
No filter or selection is applied on the signals collected on the FE boards, and the entire data stream of TDC hits is serialized with the GBTx-FPGA protocol \cite{gbtx} to SFP+ transceivers, for transmission over $5~\Gbps$ optical links to a back-end (BE) board.
No trigger signal is required to drive the data acquisition as all signals are forwarded to a BE board. 
The data stream is thus continuous and asynchronous as all signals received by the FE boards are digitized and transmitted to the optical links as soon as they are made available. 
Both optical links are collected into the BE board, a Xilinx KCU1500 evaluation board equipped with a Kintex UltraScale XKCU115 FPGA, hosted in a Dell PowerEdge R730 server.
Data from each link is deserialized by the GBTx-FPGA protocol.
The clock is recovered directly from the received data by the CDR (Clock Data Recovery) of the fast transceivers on the FPGA.
The stream from each link is then processed independently by an algorithm implemented on FPGA performing the reconstruction of the trajectory compatible with the passage of muons through each chamber.
The reconstruction algorithm~\cite{migliorini2021muon} implements a logic combining Neural Networks and analytical relations, identifying signals compatible with the local muon tracks (referred to as track \emph{segments}) and extracting the relevant information: crossing position and angle, and time of passage of the muon inside the chamber. 
All resulting parameters are merged with the original stream of hits, enriching the data stream with higher-quality information, and are not directly used to filter or select data at this stage. 
Data from both lanes are subsequently gear-boxed into a single data stream and sent to a Direct Memory Access (DMA) engine via the Advanced eXtensible Interface (AXI) stream protocol over PCIe Gen 3 bus. 
A FIFO is used as a data buffering stage to optimize the DMA throughput by defining the optimal size of the DMA data transfers.
The Xilinx AXI-DMA engine~\cite{xilinx_dma} is used for the transmission of the data stream to the server memory. 
The throughput achieved in the DMA transfers exceeds the one currently measured in the online data processing discussed in the following Section and thus does not constitute a bottleneck for the DAQ scheme described in this paper.
Overall, the FPGA resources used for the DMA and GBTx deserialization of all input links are below 10\%. 
A more detailed discussion of the resources required for the online segment reconstruction is provided in~\cite{migliorini2021muon}.
A block diagram of the logic implemented in the KCU1500 firmware is presented in Fig.~\ref{fig:kcufw}. 

\begin{figure}[thb]
    \centering
    \includegraphics[width=\linewidth]{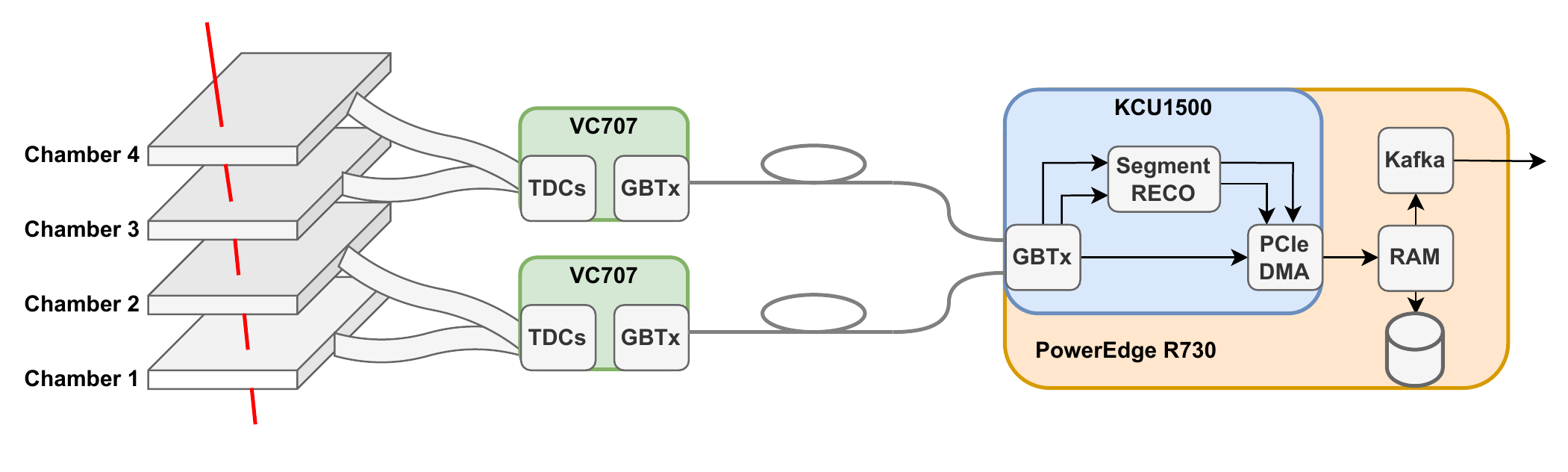}
    \caption{Simplified diagram representing the trigger-less readout chain installed in the INFN Legnaro laboratory. Left: a muon crossing the four detectors. Center: two FE boards implementing the TDC and GBTx serialization. Right: the readout server hosting the BE board where the online segment reconstruction and the DMA readout is performed.}
    \label{fig:readout}
\end{figure}

From here, the data stream can be stored locally or dispatched to a remote online processing system via an event streaming platform, described in the next section.
An overview of the readout chain used for the readout of the testbed experimental setup is presented in Fig.~\ref{fig:readout}. 

\section{Online Processing Pipeline}\label{sec:online}

\begin{figure}[thb]
    \centering
    \includegraphics[width=\linewidth]{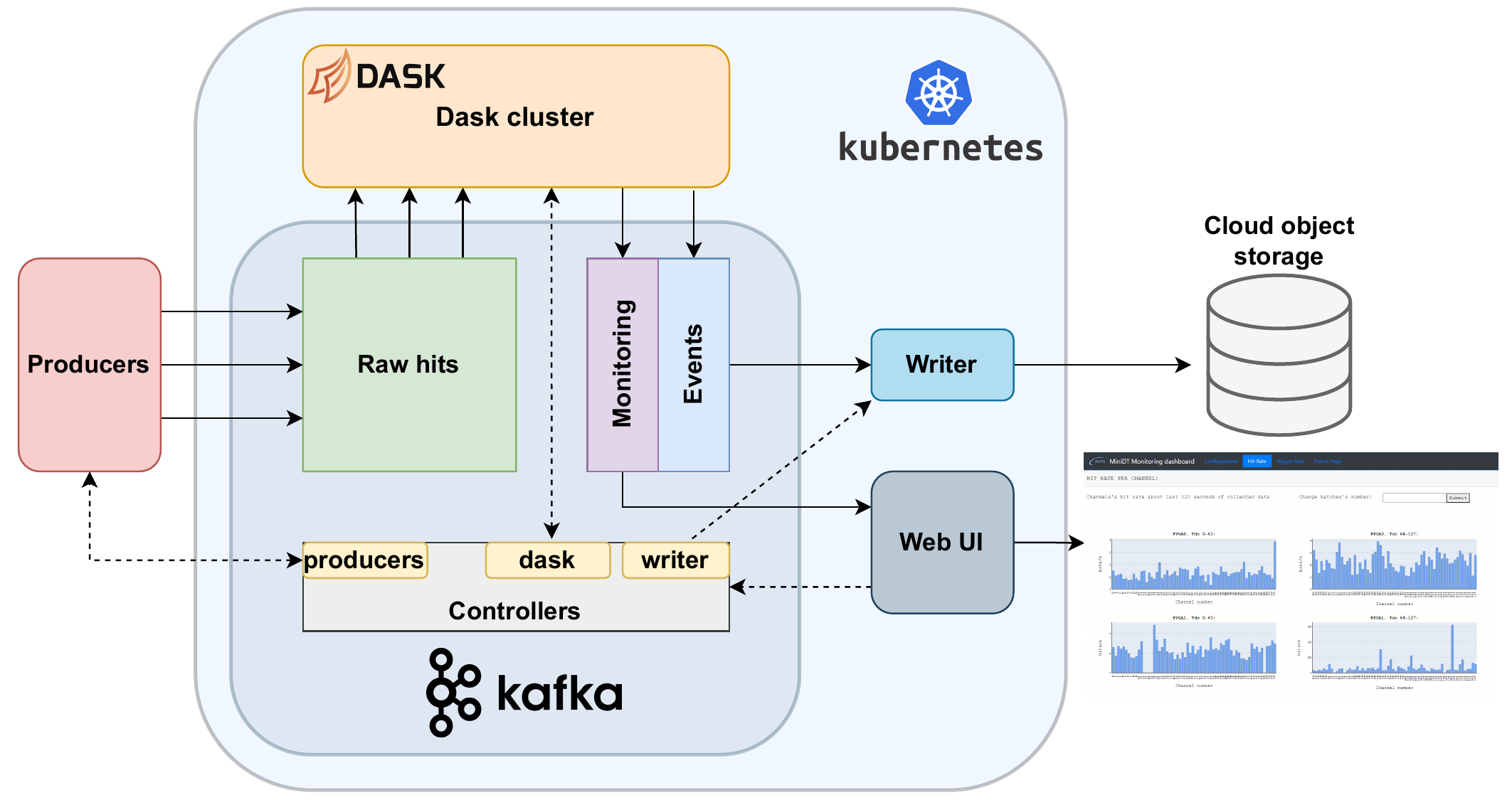}
    \caption{Schematic representation of the cluster architecture. The Kubernetes cluster acts as the orchestrator of a number of pods where the underlying processing components are hosted. The data stream generated by the DMA readout (left) is produced to a dedicated topic ("Raw Hits") of the Kafka cluster. The data is processed by the Dask cluster and the resulting products are further buffered in the Kafka cluster for visualization and storage.}
    \label{fig:cluster}
\end{figure}

The flow of TDC hits and reconstructed segments collected by the asynchronous readout system produces a continuous data stream that is processed online on a distributed computing infrastructure.
In any real-time application concerning the collection of a continuous data stream from several sources, the signals produced by a common origin may be collected in an arbitrary ordering, as a number of different delays can affect the propagation of signals from each source.
This is also true in a generic experimental setup, as the hits produced by the passage of a particle may be collected by several sub-detectors with different delays due to the length of the cables, and the delays induced by the inner logic implemented in the FE readout boards.
For these reasons, a data buffer layer is added to the online processing system to enable the collection of all hits produced by coherent signals. 
In the context of the testbed, this buffering time allows to collect and process all hits produced by the same muon in different chambers, accounting for both the internal and external sources of delay among coherent hits, such as the different electron clouds' drift times, and the different cable length per chamber, respectively.
The proposed DAQ scheme is designed to sustain a wide range of data throughput by relying on horizontally scalable tools to move data from the input sources to the computing elements and perform the distributed processing of the data stream.
The horizontal scalability of the system is a key feature of the proposed scheme.
Under this assumption, an increasing number of computing resources can be exploited to accommodate for a high number of input sources and/or large data rates.
A strong focus has been aimed at designing a system that can be easily provisioned and scaled, to allow for simple deployment on many different computing infrastructures.
All online processing components are managed using the Kubernetes orchestrator, allowing to easily scale the applications and deploy the system in any cloud or bare-metal computing cluster. 
The Kubernetes cluster used to test a first end-to-end implementation of this DAQ solution is deployed on INFN-Cloud~\cite{infncloud} infrastructure using CloudVeneto~\cite{cloudveneto} as a backbone. 
The deployed cluster is composed of 5 Virtual Machines, each with 4 vCPU, 16~GB of RAM, and 25~GB of disk. 
The amount of resources allocated is chosen to represent a deliberately small-scale computational cluster, whose overall resources are less than the ones typically available in any single node of a particle physics experiment trigger farm.
An external object storage service, accessible via the S3 interface, is used to store the products of the online data processing.
In the Kubernetes cluster, an ingress controller is installed to redirect the traffic from the public endpoint to the selected service or pod running inside the cluster. 
In this cluster, NGINX \cite{nginxswebsite} is adopted as the ingress controller. 
This allows to protect with certificates the applications exposed outside the cluster, such as the user interface used to control the DAQ. 
The role of each component in the proposed DAQ scheme is described in the following sections.
An overview of the system is presented in Fig. \ref{fig:cluster}.

\subsection{Data buffer and message queue}
 
\begin{figure}[htb]
    \centering
    \includegraphics[width=.8\linewidth]{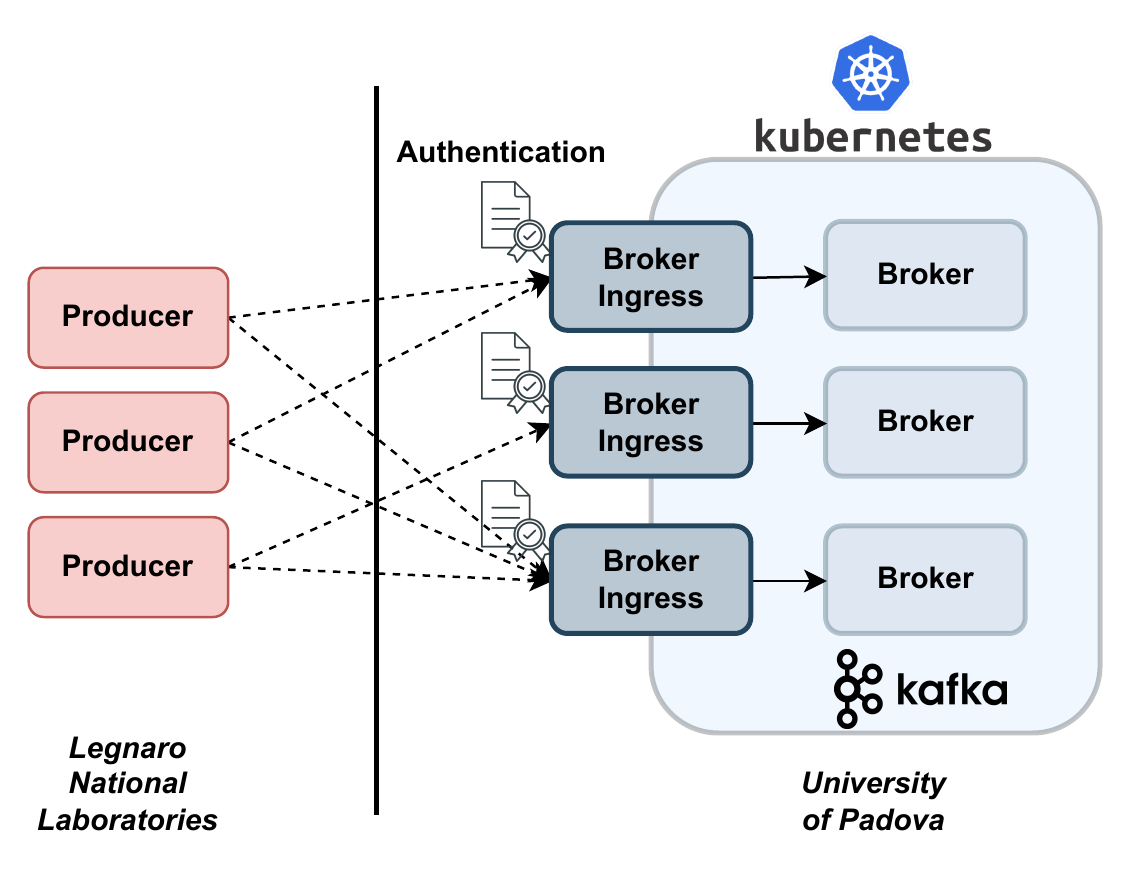}
    \caption{Schematic representation of the communication between the producers in the readout server and the brokers. The producers are physically deployed in the Legnaro National Laboratories, while the Kubernetes cluster is instantiated on resources hosted in Padova University.}
    \label{fig:kafka_lnl}
\end{figure}

Each readout unit can either write the hit stream to the a local disk or send it to a remote data buffer and messaging system. 
The latter is implemented with Apache Kafka~\cite{kafka}, a distributed event streaming platform based on a pub-sub messaging model. 
In this schema, the readout unit will be a data producer, publishing data on a specific topic, hosted in several partitions on the remote brokers. 
With reference to Fig.~\ref{fig:cluster}, the pool of readout units is named \textit{Producers} and the data stream produced by any number of BE boards is written in a topic.
The topic will thus play the role of the buffer for the data stream. 
Messages in this topic are continuously read by a set of consumers from the processing engine. 
The producer/consumer architecture enables to decouple the writing and reading processes, allowing to deal with the asynchronous nature of the readout system without inducing back-pressure to the earlier stages of the DAQ chain.
The distributed nature of the system, where the data is partitioned across multiple Kafka servers, enables the system to scale out easily, as more partitions and brokers can be added to increase the throughput and to cope with a larger number of data producers. 
Some other topics are defined in the very same Kafka cluster to deal with the results of data processing, data visualization, and data storage.
The Kafka cluster is deployed using Strimzi, a project providing container images and operators for running Apache Kafka on Kubernetes\cite{strimziswebsite}. 
Strimzi allows exposing the Kafka brokers outside the Kubernetes cluster, using the NGINX~\cite{nginxswebsite} ingress controller. 
This is especially required as the Kubernetes and the readout server are deployed in two separated networks.
The experimental setup, including the chambers and BE server, is installed in the Legnaro INFN National Laboratories, while the computing resources of CloudVeneto are physically hosted in the Physics and Astronomy Department of the University of Padova.
The two sites are located approximately $10\,\mathrm{Km}$ afar, and are hosted on a different local network. 
The Kafka clients outside the cluster (i.e. all producers as the readout server) communicate with the Kafka cluster via authentications, using a TLS certificate. 
This makes the system extremely generalizable as the same computing infrastructure could be deployed on any computing infrastructure, including public cloud computing providers such as AWS,  Google Cloud, and Azure.
A sketch of the Kafka deployment in the Kubernetes cluster is reported in Fig. \ref{fig:kafka_lnl} to highlight the aforementioned ingress and authentication.

The performances of the data brokerage component of the pipeline have been measured by generating arbitrarily large rates of synthetic hits at the level of the readout server. 
The hits are then injected into the Kafka cluster using the Kafka producer API.
%
%
Multiple Kafka producers can be used to publish data simultaneously to the same set of brokers, thus simulating a system where the data stream produced by several independent FE devices is collected and processed by a number of BE boards, to be eventually processed by the distributed computing system.
%
%
Kafka producers can be fine-tuned to optimize their performance based on the application requirements, e.g. minimize data transmission latency or maximize throughput. 
In this work, it is chosen to optimize the throughput by grouping messages in batches and compressing them before the transmission.
Combining multiple producers, an overall maximum throughput exceeding 1.7~\GBs has been measured. 
In this test, the throughput of each producer is set at 230~\MBs to mimic an arbitrarily expected throughput produced by a BE board under stress conditions. 
This would represent a continuous rate of about 25 million hits per second (corresponding to roughly 1.5 million muons traversing the testbed each second). %
All tests have been carried out on a 10~\Gbps network shared with all other users of the CloudVeneto computing infrastructure. 

\begin{figure}[htb]
    \centering
    \includegraphics[width=.8\linewidth]{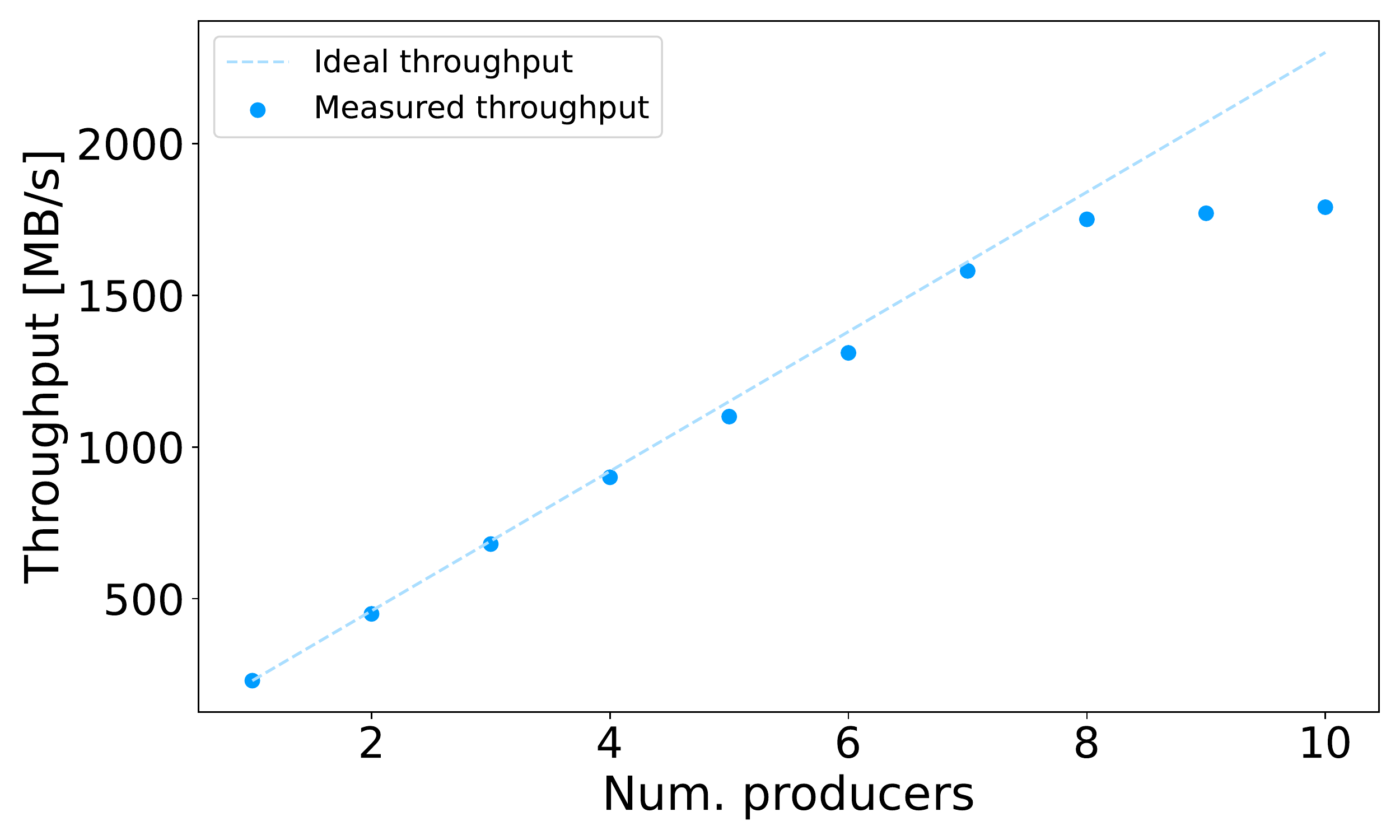}
    \caption{Measured data throughput as produced to the Kafka cluster as a function of the number of producers. Each Kafka producer is set to publish data to a fixed $230~\MBs$ throughput.}
    \label{fig:kafka_perf}
\end{figure}

An almost perfect throughput linearity has been observed with the number of producers up to 8 instances, before experiencing saturation due to networking limitations. 
The results are shown in Figure~\ref{fig:kafka_perf}.

\subsection{Processing engine}
 
Having access to the entire stream of data produced at the detectors' FEs provides the possibility for multiple levels of data processing, ranging from coarse measurements directly based on raw data, up to executing sophisticated high-level algorithms.
The trigger-less nature of the system allows having a continuous and unbiased monitoring of the detector status and data quality without the limitation induced by sampling only a small fraction of the data associated with dedicated trigger logics. 
This allows to perform data quality monitoring on the entirety of the data produced by the detectors, and to measure unbiased quantities (e.g., the overall background noise affecting the detector), typically inaccessible.
On a higher level, the goal of the processing engine may be targeted towards building high-level abstractions for the data collected, such as merging data into events or reconstructing the particle trajectories, by running simple algorithms and computing aggregations to combine all consistent data in a single structure. 
Both these processing layers are implemented in the context of this demonstrator using Dask~\cite{dask}, a Python library for parallel and distributed computing.
In the context of the testbed, intermediate-level data structures (events) are reconstructed as the set of all hits produced by the passage of a muon. 
Thanks to the local reconstruction performed in the BE board, the muons' passage times associated with the reconstructed segments can be used to cluster all hits compatible with each muon crossing time.
All hits associated with the same muon are finally grouped into a single event structure by creating a list of raw hits along with the complete information from the segment online reconstruction. 
Due to the structured nature of the data stream, the hits' batches are interpreted as DataFrames: structured datasets distributed across multiple nodes of the processing cluster.
Column-wise operations can be performed to produce a high-level reconstruction of the features, and events are produced using standard DataFrame operations. 

\begin{figure}[thb]
    \centering
    \includegraphics[width=\linewidth]{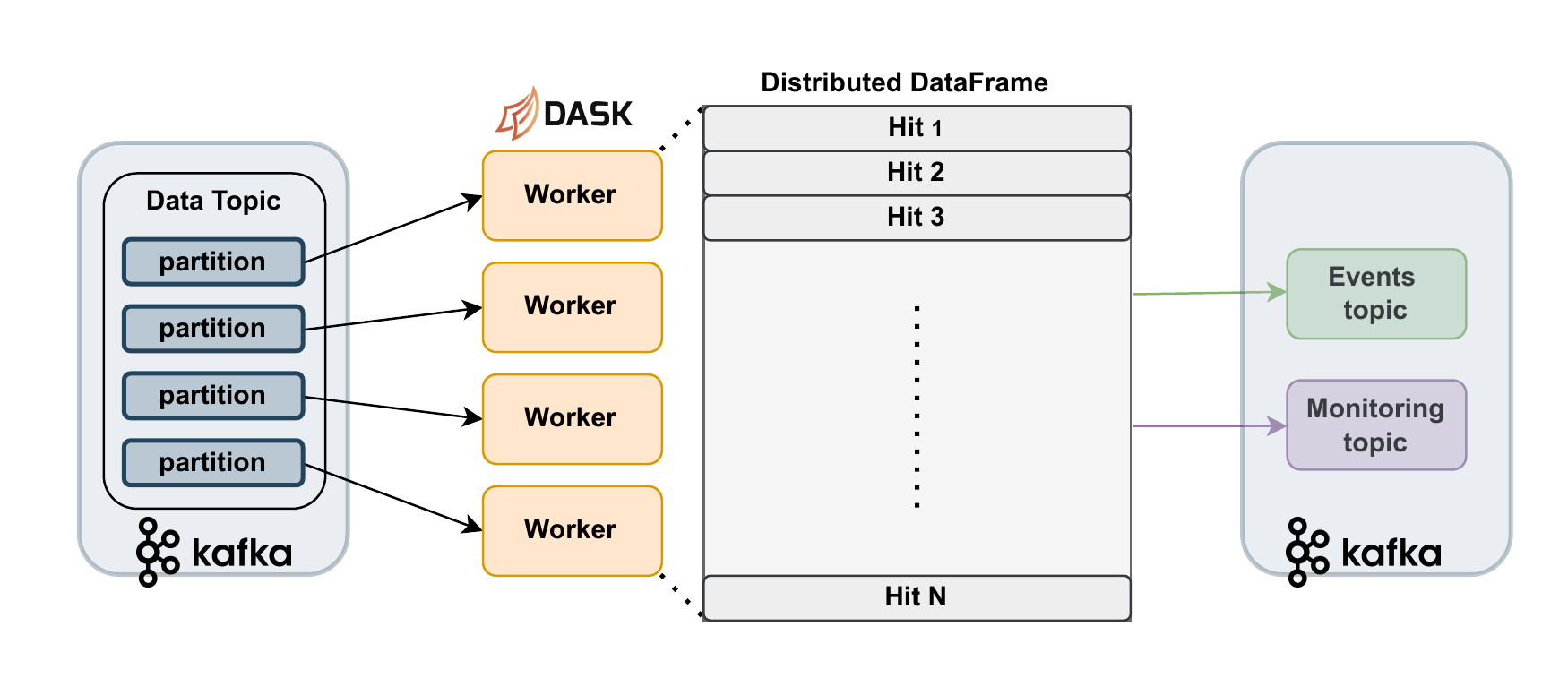}
    \caption{Schematic representation of the distributed data processing. Each Dask worker reads from one or more partitions. A Dask DataFrame is created from the batch of hits and used to compute monitoring quantities and to group hits into events, before publishing the results to the related Kafka topics.}
    \label{fig:dask}
\end{figure}

A Dask cluster is instantiated on the CloudVeneto computing infrastructure via the Kubernetes orchestrator.
Dask workers periodically read Kafka partitions to collect raw detector hits and to create a Dask distributed DataFrame, in batches of 4 seconds each. 
It should be noted that the batch duration (either based on time or data size) is a parameter that can be adjusted in software depending on a number of factors, e.g. the expected data input rate and the processing time. 
Each worker fetches hits from one or more topic partitions and stores them in its local memory. 
With this procedure, the content of each topic partition relative to the current batch is transformed into a DataFrame partition, where each block is located in one worker.
From this DataFrame, all relevant quantities are computed using the Dask DataFrame APIs and a set of Python extensions. 
Upon completion of the batch processing, events and data monitoring quantities are written in parallel from each worker back to two new Kafka topics, for further processing and visualization. 
Low-level aggregations of the raw hits are organized into lists and published to a data-monitoring topic; events are instead redirected to a dedicated event topic for storage.
The Dask processing system is illustrated in Fig. \ref{fig:dask}. 

\begin{figure}[htb]
    \centering
    \includegraphics[width=.8\linewidth]{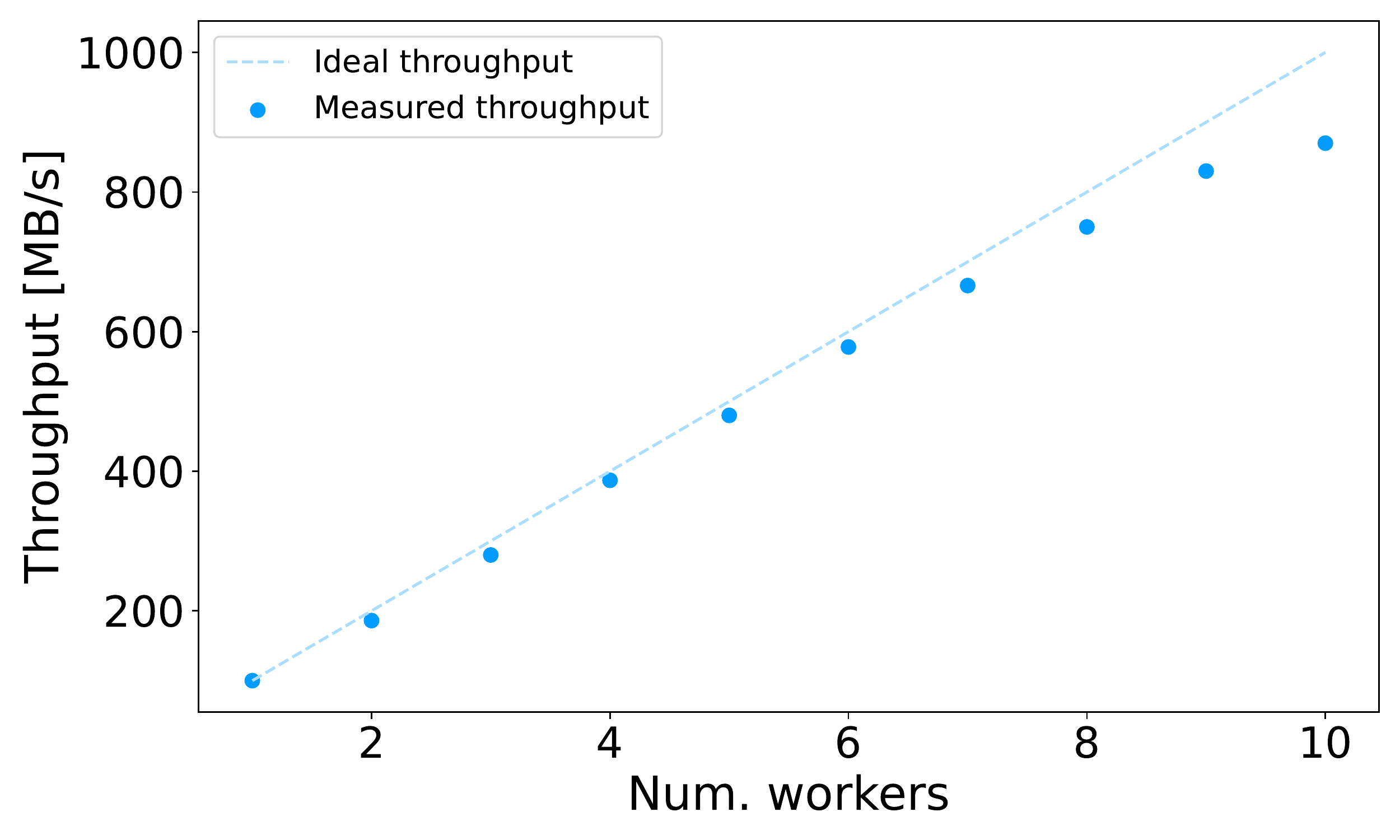}
    \caption{Measured throughput of the data processed by the Dask cluster as a function of the number of allocated workers. Each Dask worker is limited to 1 vCPU and 3~\GB of RAM.}
    \label{fig:dask_perf}
\end{figure}

The performances of the data processing component of the pipeline are deeply dependent on the type and complexity of processing that is carried out. 
It is therefore not trivial to define an unambiguous figure of merit to measure the performance of the Dask processing.
In this work, a simple task is defined to benchmark the processing framework using the allocated computing resources. 
This includes the unpacking of binary-formatted TDC hits, the creation of the distributed DataFrame across several partitions, the computation of the channel hit rate and chambers occupancy plots, and the clustering of all hits associated with a signal generated from the online reconstruction algorithm performed in the BE board. 
The number of Kafka partitions is set to match the number of Dask workers.
A Dask cluster composed of up to 10 workers, each with 1 vCPU and 3~\GB of RAM, and deployed on Kubernetes, is capable of continuously processing data at a rate of 75 million hits per second.
While it is clear that more complex and time-demanding processing, such as the full reconstruction of a large number of particles in complex collision environments, would reduce the processing rate, the horizontal scalability provided by the Dask distributed processing framework can be exploited, and more workers can be added to share the processing and adjust to sustain the throughput. 
The linearity of the processed data throughput with the number of workers has been tested by allocating a variable number of Dask workers, each with the limited resources of 1 vCPU and 3~\GB of RAM.
The results are shown in Figure~\ref{fig:dask_perf}, where excellent linearity can be observed up to about 800~\MBs.
A small deviation from linearity is observed at high throughput, due to operations involving the exchange of data across workers (data shuffling).
This can be solved with more efficient programming of the online data processing, to better exploit the parallel nature of the system.

\subsection{Controllers and Web User Interface}~\label{sec:control_and_webui}
A web User Interface (UI), developed using Plotly\cite{plotly}, allows to monitor and control each component of the DAQ chain, including issuing the start and stop of a data acquisition run. 
Via the web UI it is also possible to provide different configurations to the readout unit, such as enabling the masking of selected channels of a chamber, and setting and modifying the address and topics of the Kafka brokers where the hits are produced. 
These instructions are exchanged across the web UI and the readout server, once more via the Kafka message brokerage cluster used for the buffering of the data stream. 
For instance, as the \textit{Start Run} button is pressed on the web UI, a specific message is produced to a controller topic and read by a consumer running on the readout server. 
This will trigger the start of the data acquisition targeting the specified brokers and taking into account the set of options stored in a dedicated topic. 
At the start of a run, a message containing the run's metadata is also produced and consumed by the writer controller.
The latter automatically creates a bucket in the object storage, which will be used to store the run's reconstructed events for long-term data preservation. 
Another service, whose controls are integrated into the web UI, is in charge of controlling the Dask application.
Users can thus control its status and change the relevant parameters remotely in case of necessity. 
The status of the Dask scheduler can be continuously monitored during the execution by the dedicated dashboard provided by Dask, and exposed through a Kubernetes service.

A view of the web UI is finally devoted to the data visualization of all detector-related quantities produced to the monitoring topic, such as the channel occupancy and online segment reconstruction rate. 
The histograms and time series shown in the web UI are built from the entirety of the hits processed by the trigger-less readout system, and thus allow the live monitoring of all data collected by the detectors, with no filters due to online trigger selection.
A monitoring infrastructure for the status of the nodes of the cluster is also integrated into the web UI.
Metrics from all computing nodes hosting the cluster and containers are collected in a Prometheus~\cite{prometheuswebsite} database and visualized using Grafana dashboards~\cite{Web:Grafana:Docs}. 
The dashboard can be used to monitor parameters such as disk, CPU, and RAM usage, as well as to set alarms that may be raised in the case of anomalies in the containers and host resource utilization.

\section{Conclusions}

In this work, a proposal for an online processing scheme for trigger-less data streams has been presented.
The use of modern computing infrastructures allows performing the fast processing of the data produced by the detectors with little focus on hardware and no custom computing and networking requirements.
This can be achieved throughout the implementation of horizontally-scalable solutions based on distributed processing frameworks, such as Dask, and event-streaming platforms such as Apache Kafka.
The simple deployment of such a scheme is ensured by the choice of managing the whole processing pipeline with the Kubernetes orchestrator.
This guarantees the simple implementation, scalability, and portability of the solution in any computing infrastructure, potentially including public cloud providers.
The end-to-end implementation of this DAQ scheme described in this work, based on a set of gaseous detectors used as the experimental testbed for development, includes low- and intermediate-level processing of the data stream produced by the simple experimental setup.
This is set to exemplify a number of common use cases for the data quality monitoring of the detector and the reconstruction of events.

At an even higher level, it can be foreseen to implement more sophisticated algorithms to perform the detection of anomalies in the streaming data.
In the case of low-level data, such as the signals collected by the detectors' front-end, this schema can be put to use as an automated data quality monitoring, which can be used to identify striking anomalies and raise alerts to the data-taking crew.
A direction of future development can be also foreseen for the application of anomaly detection to the reconstructed features, where events identified as anomalies could be stored for later inspection; if deemed interesting and valuable to study, dedicated selections may then be implemented to accumulate statistically sound datasets of all events identified as anomalies.

The performances of the proposed schema, albeit preliminary at this stage of development, are encouraging and are achieved with very limited computing resources.

Future developments of the proposed setup will be focused on the optimization of the processing performance and more thorough measurements of its scalability, including the exploitation of accelerators for distributed processing.

\bibliography{mybibliography}

\begin{thebibliography}{10}
\expandafter\ifx\csname url\endcsname\relax
  \def\url#1{\texttt{#1}}\fi
\expandafter\ifx\csname urlprefix\endcsname\relax\def\urlprefix{URL }\fi
\expandafter\ifx\csname href\endcsname\relax
  \def\href#1#2{#2} \def\path#1{#1}\fi

\bibitem{cms_daq}
{The CMS Collaboration}, {The TriDAS project. Technical design report, Vol. 2:
  Data acquisition and high-level trigger}, CERN-LHCC-2002-026 (2002).

\bibitem{CMS:2016ngn}
V.~Khachatryan, et~al., {The CMS trigger system}, JINST 12~(01) (2017) P01020.
\newblock \href {http://arxiv.org/abs/1609.02366} {\path{arXiv:1609.02366}},
  \href {https://doi.org/10.1088/1748-0221/12/01/P01020}
  {\path{doi:10.1088/1748-0221/12/01/P01020}}.

\bibitem{Bawej:2014atd}
T.~Bawej, et~al., {The New CMS DAQ System for Run-2 of the LHC} (2014)
  7097437\href {https://doi.org/10.1109/RTC.2014.7097437}
  {\path{doi:10.1109/RTC.2014.7097437}}.

\bibitem{atlas_daq}
{ATLAS high-level trigger, data acquisition and controls: Technical design
  report}, CERN-LHCC-2003-022, ATLAS-TRD-016 (2003).

\bibitem{ATLASTDAQ:2016pov}
{The ATLAS TDAQ Collaboration}, {The ATLAS Data Acquisition and High Level
  Trigger system}, JINST 11~(06) (2016) P06008.
\newblock \href {https://doi.org/10.1088/1748-0221/11/06/P06008}
  {\path{doi:10.1088/1748-0221/11/06/P06008}}.

\bibitem{lhcb_daq}
{The LHCb Collaboration}, {LHCb Trigger and Online Upgrade Technical Design
  Report}, CERN-LHCC-2014-016, LHCB-TDR-016 (2014).

\bibitem{cms}
{The CMS Collaboration}, {The CMS experiment at the CERN LHC. The Compact Muon
  Solenoid experiment}, JINST 3 (2008) S08004. 361 p.
\newblock \href {https://doi.org/10.1088/1748-0221/3/08/S08004}
  {\path{doi:10.1088/1748-0221/3/08/S08004}}.

\bibitem{Lee:2018pag}
L.~Lee, C.~Ohm, A.~Soffer, T.-T. Yu, {Collider Searches for Long-Lived
  Particles Beyond the Standard Model}, Prog. Part. Nucl. Phys. 106 (2019)
  210--255.
\newblock \href {http://arxiv.org/abs/1810.12602} {\path{arXiv:1810.12602}},
  \href {https://doi.org/10.1016/j.ppnp.2019.02.006}
  {\path{doi:10.1016/j.ppnp.2019.02.006}}.

\bibitem{lhcb_turbo}
S.~Benson, V.~V. Gligorov, M.~A. Vesterinen, M.~Williams, {The LHCb Turbo
  Stream}, J. Phys. Conf. Ser. 664~(8) (2015) 082004.
\newblock \href {https://doi.org/10.1088/1742-6596/664/8/082004}
  {\path{doi:10.1088/1742-6596/664/8/082004}}.

\bibitem{xenon1t}
E.~Aprile, et~al., {The XENON1T Data Acquisition System}, JINST 14~(07) (2019)
  P07016.
\newblock \href {http://arxiv.org/abs/1906.00819} {\path{arXiv:1906.00819}},
  \href {https://doi.org/10.1088/1748-0221/14/07/P07016}
  {\path{doi:10.1088/1748-0221/14/07/P07016}}.

\bibitem{Kuberneteswebsite}
Kubernetes, available at \url{https://kubernetes.io/}.

\bibitem{CMS:2009lnj}
S.~Chatrchyan, et~al., {Performance of the CMS Drift Tube Chambers with Cosmic
  Rays}, JINST 5 (2010) T03015.
\newblock \href {http://arxiv.org/abs/0911.4855} {\path{arXiv:0911.4855}},
  \href {https://doi.org/10.1088/1748-0221/5/03/T03015}
  {\path{doi:10.1088/1748-0221/5/03/T03015}}.

\bibitem{lemma}
N.~Amapane, et~al., {Study of muon pair production from positron annihilation
  at threshold energy}, JINST 15~(01) (2020) P01036.
\newblock \href {http://arxiv.org/abs/1909.13716} {\path{arXiv:1909.13716}},
  \href {https://doi.org/10.1088/1748-0221/15/01/P01036}
  {\path{doi:10.1088/1748-0221/15/01/P01036}}.

\bibitem{cms_mad}
F.~Gonella, M.~Pegoraro, {'The MAD', a full custom ASIC for the CMS barrel muon
  chambers front end electronics}, in: {7th Workshop on Electronics for LHC
  Experiments}, 2001, pp. 204--208.

\bibitem{cms_tcds}
J.~Hegeman, et~al., {The CMS Timing and Control Distribution System}, in: {2015
  IEEE Nuclear Science Symposium and Medical Imaging Conference}, 2016, p.
  7581984.
\newblock \href {https://doi.org/10.1109/NSSMIC.2015.7581984}
  {\path{doi:10.1109/NSSMIC.2015.7581984}}.

\bibitem{gbtx}
S.~Baron, J.~P. Cachemiche, F.~Marin, P.~Moreira, C.~Soos, {Implementing the
  GBT data transmission protocol in FPGAs}, in: {Topical Workshop on
  Electronics for Particle Physics}, CERN, 2009.
\newblock \href {https://doi.org/10.5170/CERN-2009-006.631}
  {\path{doi:10.5170/CERN-2009-006.631}}.

\bibitem{migliorini2021muon}
M.~Migliorini, J.~Pazzini, A.~Triossi, M.~Zanetti, A.~Zucchetta, Muon trigger
  with fast neural networks on fpga, a demonstrator (2021).
\newblock \href {http://arxiv.org/abs/2105.04428} {\path{arXiv:2105.04428}}.

\bibitem{xilinx_dma}
{Xilinx, Inc.}, {DMA}/bridge subsystem for {PCI} express v4.1, available at
  \url{https://www.xilinx.com/support/documentation/ip_documentation/xdma/v4_1/pg195-pcie-dma.pdf}.

\bibitem{infncloud}
The INFN Cloud, available at \url{https://www.cloud.infn.it/}.

\bibitem{cloudveneto}
P.~Andreetto, F.~Chiarello, F.~Costa, A.~Crescente, S.~Fantinel, F.~Fanzago,
  E.~Konomi, P.~Mazzon, M.~Menguzzato, M.~Segatta, G.~Sella, M.~Sgaravatto,
  S.~Traldi, M.~Verlato, L.~Zangrando, Merging openstack-based private clouds:
  the case of cloudveneto.it, EPJ Web of Conferences 214 (2019) 07010.
\newblock \href {https://doi.org/10.1051/epjconf/201921407010}
  {\path{doi:10.1051/epjconf/201921407010}}.

\bibitem{nginxswebsite}
NGINX ingress controller, available at
  \url{https://kubernetes.github.io/ingress-nginx/}.

\bibitem{kafka}
Apache Kafka, available at \url{https://kafka.apache.org/}.

\bibitem{strimziswebsite}
Strimzi, available at \url{https://strimzi.io/}.

\bibitem{dask}
{Dask Development Team}, Dask: Library for dynamic task scheduling, available
  at \url{https://dask.org}.

\bibitem{plotly}
{Plotly Technologies Inc.}, Collaborative data science, available at
  \url{https://plot.ly}.

\bibitem{prometheuswebsite}
Prometheus, available at \url{https://prometheus.io/}.

\bibitem{Web:Grafana:Docs}
Grafana, available at \url{https://grafana.com/docs/}.

\end{thebibliography}

\end{document}